\documentclass[aps,prl,showpacs]{revtex4}
\usepackage{amssymb}

\usepackage{amsmath}
\usepackage{graphicx}
\usepackage{bm}

\begin{document}

\title{A systematic study
on the exact solution of the position dependent mass Schr\"{o}dinger equation%
}
\date{\today}
\author{Ramazan Ko\c{c}}
\email{koc@gantep.edu.tr}
\affiliation{Department of Physics, Faculty of Engineering University of Gaziantep, 27310
Gaziantep, Turkey}
\author{Mehmet Koca}
\email{kocam@squ.edu.om}
\affiliation{Department of Physics, College of Science, Sultan Qaboos University, PO Box
36, Al-Khod 123, Muscat, Sultanate of Oman}
\
\begin{abstract}
An algebraic method of constructing potentials for which the Schr\"{o}dinger
equation with position dependent mass can be solved exactly is presented. A
general form of the generators of su(1,1) algebra has been employed with a
unified approach to the problem. Our systematic approach reproduces a number
of earlier results and also leads to some novelties. We show that the
solutions of the Schr\"{o}dinger equation with position dependent mass are
free from the choice of parameters for position dependent mass. Two classes
of potentials are constructed that include almost all exactly solvable
potentials.
\end{abstract}

\maketitle

\section{Introduction}

The study of position dependent mass (PDM) Schr\"{o}dinger equation has
recently attracted some interest\cite{roy, milan} arising from the study of
electronic properties of semiconductors, liquid crystals, quantum dots, the
recent progress of crystal-growth techniques for production of non-uniform
semiconductor specimen in which carrier effective mass depends on position%
\cite{serra}. It is obvious that the study of PDM Schr\"{o}dinger equation
has considerable impact on condensed matter physics as well as related
fields of the physics.

Exact solvability of the Schr\"{o}dinger equation with constant mass has
been the main interest since the early days of quantum mechanics\cite{levai}%
. It has been solved exactly for a large number of potentials by employing
various techniques. In fact, for exactly solvable potentials its general
solution can be obtained in terms of some special functions by transforming
the original Schr\"{o}dinger equation into the second order differential
equation . Systematic studies of these transformations have been given in%
\cite{natan} regarding the confluent hypergeometric and hypergeometric
functions. The relations between the algebraic technique and the special
function theory have been discussed in\cite{cordero}. Recently various
approaches have been presented in a unified way and a number of earlier
results have been reproduced\cite{levai}. In the present work we use the Lie
algebraic technique to construct the Hamiltonian for the PDM Schr\"{o}dinger
equation and obtain the solutions in terms of the special functions.

In the PDM Schr\"{o}dinger equation the mass and momentum operator no longer
commute. The general expression for the kinetic energy operator have been
introduced by von Roos\cite{roos}:%
\begin{equation}
T=\frac{1}{4}\left( m^{\eta }\mathbf{p}m^{\varepsilon }\mathbf{p}m^{\rho
}+m^{\rho }\mathbf{p}m^{\varepsilon }\mathbf{p}m^{\eta }\right)  \label{eq:1}
\end{equation}%
where $\eta +\varepsilon +\rho =-1$ is a constraint. One of the problem is
the choice of parameters\cite{dutra, dekar}. In our approach we obtain exact
solution of the PDM Schr\"{o}dinger equation without any particular choice
which leads to a general solution where the choice of the parameters
distinguishes the physical systems.

One can display a number of fruitful applications of the Lie algebraic
technique, in particular, in atomic and nuclear physics and other fields of
the physics. Our task here is to obtain the exact solution of the PDM Schr%
\"{o}dinger equation by the use of the $su(1,1)$ algebra technique.

The paper is organized as follows. In section 2 we present a general
Hamiltonian by using $su(1,1)$ algebra and we discuss its relation with the
PDM Schr\"{o}dinger equation. We obtain a general expression for the
potential. In section 3 we describe the application of the $su(1,1)$ algebra
to obtain Coulomb, harmonic oscillator and Morse family potentials. In
section 4 we construct hyperbolic and trigonometric potentials. Finally we
discuss our results in section 5.

\section{Structure of the $su(1,1)$ Lie algebra and PDM Schr\"{o}dinger
equation}

Lie algebraic technique is suitable to study PDM Schr\"{o}dinger equation,
because they contain a first-derivative term. The $su(1,1)$ Lie algebra is
described by the commutation relations,%
\begin{equation}
\left[ J_{+},J_{-}\right] =-2J_{0},\quad \left[ J_{0},J_{\pm }\right] =\pm
J_{\pm }.  \label{eq:2}
\end{equation}%
Casimir operator of this structure is given by%
\begin{equation}
J^{2}=-J_{\pm }J_{\mp }+J_{0}^{2}\mp J_{0}.  \label{eq:3}
\end{equation}%
The eigenstate of $J^{2}$ and $J_{0}$ can be denoted by $|jN>$ where%
\begin{equation}
J^{2}|jN>=j(j+1)|jN>,\quad J_{0}|jN>=N|jN>  \label{eq:4}
\end{equation}%
while the allowed values of $N$ are%
\begin{equation}
N=-j,-j+1,-j+2,\cdots =(n+j)  \label{eq:5}
\end{equation}%
where $n$ is an integer. We consider the most general form of the generators
of the algebra which introduced by Sukumar\cite{sukumar}%
\begin{eqnarray}
J_{\pm } &=&e^{\pm i\phi }\left( \pm h(x)\frac{\partial }{\partial x}\right)
\pm g(x)+f(x)J_{0}+c(x)  \notag \\
J_{0} &=&-i\frac{\partial }{\partial \phi }.  \label{eq:6}
\end{eqnarray}%
The commutation relations (\ref{eq:2}) is satisfied when the functions $%
h(x),\quad f(x)$ and $c(x)$ takes the forms%
\begin{equation}
h(x)=\frac{r}{r^{\prime }},\quad f(x)=\frac{1+ar^{2}}{1-ar^{2}},\quad c(x)=-%
\frac{br}{1-ar^{2}}  \label{eq:7}
\end{equation}%
where $r=r(x)$ and $a$ and $b$ are constants. The differential realization (%
\ref{eq:6}) can be used to derive the second order differential equations of
the orthogonal polynomials. The differential equations of these polynomials
can be expressed in terms of Casimir operator $J^{2}$:%
\begin{equation}
H=J^{2};\quad H|jN>=j(j+1)|jN>.  \label{eq:8}
\end{equation}%
Let us consider the basis function,%
\begin{equation}
|jN>=e^{-iN\phi }\Re _{jN}(x).  \label{eq:9}
\end{equation}%
Interms of the realizations (\ref{eq:6}) and with the basis (\ref{eq:9}) the
Hamiltonian (\ref{eq:8}) takes the form 
\begin{eqnarray}
H &=&\frac{r^{2}}{r^{\prime 2}}\frac{d^{2}}{dx^{2}}+\frac{r}{r^{\prime }}%
\left( 2g-\frac{2ar^{2}}{1-ar^{2}}-\frac{rr^{\prime \prime }}{r^{\prime 2}}%
\right) \frac{d}{dx}+  \notag \\
&&4g^{2}+g+\frac{rg^{\prime }}{r^{\prime }}-\frac{2g}{1-ar^{2}}-\frac{%
r(2N+br)(2aNr+b)}{(1-ar^{2})^{2}}.  \label{eq:10}
\end{eqnarray}%
Let us now turn our attention to the PDM Schr\"{o}dinger equation which can
be written as%
\begin{equation}
H^{\prime }=T+V(x),\quad H^{\prime }\psi (x)=E\psi (x)  \label{eq:12}
\end{equation}%
where $V(x)$ is the potential of the physical system and $\psi (x)$ and $E$
are eigenstates and eigenvalues of the PDM Schr\"{o}dinger equation.
Introducing the eigenfunction and momentum operator $p$%
\begin{equation}
\psi (x)=-\frac{2mr^{2}}{r^{\prime 2}}\Re (x);\quad p=-i\frac{d}{dx}
\label{eq:13}
\end{equation}%
respectively, then the position dependent mass Hamiltonian takes the form%
\begin{eqnarray}
H^{\prime } &=&\frac{r^{2}}{r^{\prime 2}}\frac{d^{2}}{dx^{2}}+\frac{r}{%
r^{\prime }}\left( 4-\frac{4rr^{\prime \prime }}{r^{\prime 2}}+\frac{%
rm^{\prime }}{r^{\prime }m}\right) \frac{d}{dx}+  \notag \\
&&2+\frac{2r}{r^{\prime 2}}\left( \frac{3rr^{\prime \prime 2}}{r^{\prime 2}}-%
\frac{rr^{\prime \prime \prime }}{r^{\prime }}-3r^{\prime \prime }\right) + 
\notag \\
&&\frac{m^{\prime }r^{2}}{mr^{\prime 2}}\left( \frac{(1+\eta )(\varepsilon
+\eta )m^{\prime }}{m}+\frac{(1-\varepsilon )m^{\prime \prime }}{2m}+\frac{%
2(r^{\prime 2}-rr^{\prime \prime })}{rr^{\prime }}\right) -  \label{eq:13x}
\\
&&\frac{2mr^{2}}{r^{\prime 2}}V(x)
\end{eqnarray}%
then comparing the (\ref{eq:13x}), (\ref{eq:8}), (\ref{eq:10}) and we obtain
the following general expression for the potential,%
\begin{eqnarray}
V(x)-E &=&  \notag \\
&&\frac{(2bN+r(b^{2}+a(4N^{2}-1)+2abNr))r^{\prime 2}}{2mr(1-ar^{2})^{2}}+ 
\notag \\
&&\frac{(j(j+1))r^{\prime 2}}{2mr^{2}}+\frac{3r^{\prime \prime ^{2}}}{%
8mr^{\prime 2}}-\frac{r^{\prime \prime \prime }}{4mr^{\prime }}+V_{m}(x)
\label{eq:14}
\end{eqnarray}%
where $V_{m}(x)$ is given by%
\begin{equation}
V_{m}(x)=\frac{1}{4m^{2}}\left( \frac{\left( 4\varepsilon (1+\eta )+(1+2\eta
)^{2}\right) m^{\prime 2}}{2m}-\varepsilon m^{\prime \prime }\right) .
\label{eq:15}
\end{equation}%
when the function $g(x)$ constrained to%
\begin{equation}
g(x)=\frac{ar^{2}-2}{ar^{2}-1}+\frac{m^{\prime }r}{2mr^{\prime }}-\frac{%
3rr^{\prime \prime }}{2r^{\prime 2}}.  \label{eq:11}
\end{equation}%
We note here that the potential reduces to the Natanzon class potentials for
the constant mass. In the following section we construct the quantum
mechanical potentials.

\section{Coulomb, Harmonic oscillator and Morse family potentials}

In order to obtain the corresponding potentials we choose $a=0,$ then the
potential (\ref{eq:14}) takes the form%
\begin{eqnarray}
V(x)-E &=&\left( \frac{b^{2}}{2}+\frac{j(j+1)}{2r^{2}}+\frac{bN}{r}\right) 
\frac{r^{\prime 2}}{m}+  \notag \\
&&\frac{3r^{\prime \prime ^{2}}}{8mr^{\prime 2}}-\frac{r^{\prime \prime
\prime }}{4mr^{\prime }}+V_{m}(x).  \label{eq:16}
\end{eqnarray}%
In the above potential the energy term on the left-hand side should be
represent by a constant term of the right-hand side. This condition can be
satisfied when%
\begin{equation}
\left( \lambda _{0}+\lambda _{1}r^{-1}+\lambda _{2}r^{-2}\right) \frac{%
r^{\prime 2}}{m}=1  \label{eq:17}
\end{equation}%
where $\lambda _{0},\quad \lambda _{1}$ and $\lambda _{2}$ are constants.
Choosing appropriate values of $\lambda _{0},\quad \lambda _{1}$ and $%
\lambda _{2}$ one can generate quantum mechanical potentials.

\subsection{Coulomb family potentials}

In order to generate Coulomb family potentials we choose $\lambda _{0}=1,$
and$\quad \lambda _{1}=$ $\lambda _{2}=0$. Solving (\ref{eq:17}) for $r$ and
substituting in to (\ref{eq:16}) we obtain the following potential%
\begin{equation}
V(x)=\frac{j(j+1)}{2u^{2}}+\frac{Ze^{2}}{2u}+U_{m}(x)  \label{eq:18}
\end{equation}%
with the eigenvalues 
\begin{equation}
E=-\frac{Z^{2}e^{4}}{2N^{2}}.  \label{eq:19}
\end{equation}%
where $u=\int_{0}^{x}\sqrt{m}dx,$ and the parameter $b$ of the potential (%
\ref{eq:16}) is defined as $\quad b=Ze^{2}/N$. The potential is isospectral
with the constant mass Schr\"{o}dinger equation. The mass dependent function
U$_{m}(x)$ is given by%
\begin{equation}
U(m)=\frac{m^{\prime }}{8m^{2}}\left( \frac{5m^{\prime }}{4m}-\frac{%
m^{\prime \prime }}{m^{\prime }}\right) +V_{m}(x)
\end{equation}

\subsection{Harmonic oscillator potential}

The harmonic oscillator potential can be generated from (\ref{eq:16}) when
we set the parameter $A=-\frac{3}{16}(1+2j)^{2}$, under the condition $%
\lambda _{1}=1/2,$ and$\quad \lambda _{0}=$ $\lambda _{2}=0$. In this case $%
r=\frac{u^{2}}{2}$, and the potential takes the form 
\begin{equation}
V=\frac{3+16j(j+1)}{8u^{2}}+\frac{b}{2}u^{2}+U_{m}(x)  \label{eq:20}
\end{equation}%
with the eigenvalues%
\begin{equation}
E=2bN  \label{eq:21}
\end{equation}

\subsection{Morse family potential}

Our last example in this class of potential is the Morse family potential.
This potential can be obtained by setting parameters $\lambda _{2}=1,$ and$%
\quad \lambda _{0}=$ $\lambda _{1}=0$. Solving (\ref{eq:17}) for $r$ we
obtain $r=e^{-\alpha u}$ and the potential takes the form%
\begin{equation}
V(x)=Nb\alpha ^{2}e^{-\alpha u}+\frac{b^{2}\alpha ^{2}}{2}e^{-2\alpha
u}+U_{m}(x)  \label{eq:22}
\end{equation}%
with the eigenvalues%
\begin{equation}
E=-\frac{\alpha ^{2}}{8}\left( (1+2j)^{2}\right)  \label{eq:23}
\end{equation}

\section{Hyperbolic and Trigonometric Potentials}

In this section we construct hyperbolic and trigonometric potentials. Some
of these potentials have important applications in condensed matter
phenomena because of its periodicity. As we mentioned before in the
potential (\ref{eq:14}) a constant term should be represented with the
energy term. We discuss below the problem for various potentials.

\subsection{P\"{o}schl-Teller family potential}

For the choice of $r=e^{-2\alpha u},\quad a=-1$ the result is \ 
\begin{subequations}
\begin{eqnarray}
V(x) &=&\frac{\alpha ^{2}}{8}((b-2N)^{2}-1)\csc h^{2}\alpha u-
\label{eq:24a} \\
&&\frac{\alpha ^{2}}{8}((b+2N)^{2}-1)\sec h^{2}\alpha u+U_{m}(x)  \notag \\
E &=&-\frac{\alpha ^{2}}{2}\left( (1+2j)^{2}\right)  \label{eq:24b}
\end{eqnarray}%
which is the P\"{o}schl-Teller potential. The function $u$ is given by 
\end{subequations}
\begin{equation}
u=\int_{0}^{x}\sqrt{m}dx.  \label{eq:25}
\end{equation}%
For the given mass term $u$ should be integrable. The trigonometric form of
the P\"{o}schl-Teller potential can be obtained by substituting $\alpha
\rightarrow i\alpha .$ In this case the potential and its eigenvalues are
given by 
\begin{subequations}
\begin{eqnarray}
V(x) &=&\frac{\alpha ^{2}}{8}((b-2N)^{2}-1)\csc ^{2}\alpha x+  \label{et:1}
\\
&&\frac{\alpha ^{2}}{8}(\left( b+2N\right) ^{2}-1)\sec ^{2}\alpha x+U_{m}(x)
\notag \\
E &=&\frac{\alpha ^{2}}{2}\left( (1+2j)^{2}\right)  \label{et:2}
\end{eqnarray}

\subsection{Generalized P\"{o}schl-Teller family potential}

In order to construct the generalized P\"{o}schl-Teller family potential we
introduce 
\end{subequations}
\begin{equation}
r=e^{-\alpha u},a=-1.  \label{eq:26}
\end{equation}%
Substituting $r$ into (\ref{eq:14}) the resulting potential and
corresponding eigenvalues read as 
\begin{subequations}
\begin{eqnarray}
V(x) &=&\frac{\alpha ^{2}}{8}(b^{2}+4N^{2}-1)\csc h^{2}\alpha u-
\label{eq:27a} \\
&&\frac{\alpha ^{2}}{2}bN\coth \alpha u\csc h\alpha u+U_{m}(x)  \notag \\
E &=&-\frac{\alpha ^{2}}{8}\left( 4A+(1+2j)^{2}\right)  \label{eq:27b}
\end{eqnarray}%
Trigonometric form of this potential can be obtained replacing $\alpha $ by $%
i\alpha .$Then the potential is given by 
\end{subequations}
\begin{subequations}
\begin{eqnarray}
V(x) &=&\frac{\alpha ^{2}}{8}(b^{2}+4N^{2}-1)\csc ^{2}\alpha u-  \label{et:3}
\\
&&\frac{\alpha ^{2}}{2}bN\cot \alpha u\csc \alpha u+U_{m}(x)  \notag \\
E &=&\frac{\alpha ^{2}}{8}\left( (1+2j)^{2}\right) .  \label{et:4}
\end{eqnarray}

\subsection{Scarf family potential}

Let us now construct another potential by substituting $r=ie^{-\alpha
u},\quad a=-1$ into equation (\ref{eq:14}). In this case we obtain PT
symmetric Scarf family potential\cite{bender}, 
\end{subequations}
\begin{subequations}
\begin{eqnarray}
V(x) &=&-\frac{\alpha ^{2}}{8}(b^{2}+4N^{2}-1)\sec h^{2}\alpha u+
\label{eq:28a} \\
&&\frac{i\alpha ^{2}}{2}bN\sec h\alpha u\tanh \alpha u+U_{m}(x)  \notag \\
E &=&-\frac{\alpha ^{2}}{8}\left( (1+2j)^{2}\right) .  \label{eq:28b}
\end{eqnarray}%
When we replace $b\rightarrow ib$ then the potential becomes Scarf family
potential. When we replace $\alpha $ by $i\alpha :$%
\end{subequations}
\begin{subequations}
\begin{eqnarray}
V(x) &=&\frac{\alpha ^{2}}{8}(b^{2}+4N^{2}-1)\sec h^{2}\alpha u+
\label{et:5} \\
&&\frac{\alpha ^{2}}{2}bN\sec \alpha u\tan \alpha u+U_{m}(x)  \notag \\
E &=&\frac{\alpha ^{2}}{8}\left( (1+2j)^{2}\right)  \label{et:6}
\end{eqnarray}%
The Scarf, PT symmetric Scarf and Generalized P\"{o}schl-Teller potentials
are isospectral potentials. The last six potentials have already constructed
by choosing r as an exponential function. This property implies that these
potentials form the same family potentials and they can be obtained from
each others by a simple coordinate transformation.

\subsection{Eckart family potential}

The Eckart family potential can be constructed by introducing $\ r=\coth 
\frac{\alpha x}{2},\quad a=-1$ The corresponding potential and eigenvalues
are given by 
\end{subequations}
\begin{subequations}
\begin{eqnarray}
V(x) &=&-\frac{\alpha ^{2}}{2}bN\coth \alpha u+  \label{eq:29a} \\
&&\frac{\alpha ^{2}}{2}(A+j(j+1))\csc h^{2}\alpha u+U_{m}(x)  \notag \\
E &=&-\frac{\alpha ^{2}}{8}(b^{2}+N^{2})  \label{eq:29b}
\end{eqnarray}%
Trigonometric form of this potential can be obtained by the choice of 
\end{subequations}
\begin{equation}
r=\cot \frac{\alpha x}{2},a=-1,b\rightarrow ib  \label{et:7}
\end{equation}%
then the potential (\ref{eq:14}) takes the form 
\begin{subequations}
\begin{eqnarray}
V(x) &=&\frac{\alpha ^{2}}{2}bN\cot \alpha u+  \label{et:8} \\
&&\frac{\alpha ^{2}}{2}(j(j+1))\csc ^{2}\alpha u+U_{m}(x)  \notag \\
E &=&-\frac{\alpha ^{2}}{8}(b^{2}-4N^{2})  \label{et:9}
\end{eqnarray}

\subsection{Hulthen family potential}

Another important potential of the quantum mechanic is the Hulthen
potential. the choice of $r=\coth \frac{\alpha x}{4},\quad a=-1$ produce the
following potential, 
\end{subequations}
\begin{subequations}
\begin{eqnarray}
V &=&\frac{(j(j+1)-bN/2)\alpha ^{2}e^{-\alpha u}}{2(1-e^{-\alpha u})}+
\label{eq:30a} \\
&&\frac{(j(j+1))\alpha ^{2}e^{-2\alpha u}}{2(1-e^{-\alpha u})^{2}}+U_{m}(x) 
\notag \\
E &=&-\frac{\alpha ^{2}}{32}(b-2N)^{2}  \label{eq:30b}
\end{eqnarray}

\subsection{Rosen-Morse family potential}

The last example in this category is the Rosen-Morse family potential. This
potential is isospectral with the Eckart family potential and can be
obtained by introducing 
\end{subequations}
\begin{equation}
r=\coth \left( \frac{\alpha x}{2}+i\frac{\pi }{4}\right) ,\quad a=-1
\label{eq:31}
\end{equation}%
Substituting (\ref{eq:31}) into (\ref{eq:14}) we obtain the following
potential with the eigenvalues $E$%
\begin{subequations}
\begin{eqnarray}
V(x) &=&-\frac{\alpha ^{2}}{2}bN\tanh \alpha u-  \label{eq:32a} \\
&&\frac{\alpha ^{2}}{2}(j(j+1))\sec h^{2}\alpha u+U_{m}(x)  \notag \\
E &=&-\frac{\alpha ^{2}}{8}(b^{2}+4N^{2})  \label{eq:32b}
\end{eqnarray}%
In order to obtain trigonometric form of the Rosen-Morse family potential we
substitute 
\end{subequations}
\begin{equation}
r=-i\cot \left( \frac{\alpha x}{2}+\frac{\pi }{4}\right) ,a=-1,b\rightarrow
ib  \label{et:10}
\end{equation}%
into (\ref{eq:14}) and we obtain the following potential 
\begin{subequations}
\begin{eqnarray}
V(x) &=&-\frac{\alpha ^{2}}{2}bN\tan \alpha u+  \label{et:11} \\
&&\frac{\alpha ^{2}}{2}(j(j+1))\sec ^{2}\alpha u+U_{m}(x)  \notag \\
E &=&-\frac{\alpha ^{2}}{8}(b^{2}-4N^{2})  \label{et:12}
\end{eqnarray}%
It is obvious that the Eckart, Hulten and Rosen-Morse family potentials can
be mapped onto each others by a simple coordinate transformation.

\section{Conclusions}

In this work we have made a systematic study to obtain the exact solution of
the PDM Schr\"{o}dinger equation within the context $su(1,1)$ algebra. We
have obtained a number of potentials some of which are already known while
the others are new. Another issue here is that the choice of the parameters $%
\rho $, $\eta $ and $\varepsilon $. It has been shown that the exact
solvability of the PDM Schr\"{o}dinger equation is independent of the these
parameters.

\end{subequations}

\end{document}